\documentclass[reprint,prl,twocolumn,superscriptaddress,showpacs,showkeys,floatfix,preprintnumbers]{revtex4-1}

\usepackage{braket}
\usepackage{xargs}
\usepackage{mathtools}
\usepackage[english]{babel}
\usepackage[utf8]{inputenc}
\usepackage{amsmath,amssymb,braket,slashed,mathrsfs}
\usepackage{pifont}
\usepackage{MnSymbol}
\usepackage{graphicx}
\usepackage{tikz}
\usetikzlibrary{shapes,arrows,positioning}
\tikzset{decision/.style={diamond, draw, fill=blue!20, text width=4.5em, text badly centered, inner sep=0pt}}
\tikzset{block/.style={rectangle, draw, fill=blue!20, text width=10em, text centered, rounded corners, minimum width=3.5cm}}
\tikzset{block1/.style={rectangle, draw, fill=blue!20, text width=18.5em, text centered, rounded corners, minimum width=3.5cm}}
\tikzset{line/.style={draw, -latex, thick}}
\usepackage{color,hyperref}
\usepackage{multirow,array}
\usepackage{physics}

\usepackage{cleveref}
\newcommand{\be}{\begin{equation}}
	\newcommand{\ee}{\end{equation}}
\newcommand{\ba}{\begin{eqnarray}}
\newcommand{\ea}{\end{eqnarray}}

\newcommand{\nn}{\nonumber}

\renewcommand{\vec}[1]{\mathbf{#1}}
\newcommand{\ud}{\mathrm{d}}

\newcommand{\innovation}{Collaborative Innovation Center of Quantum Matter, Beijing 100871, China}
\newcommand{\chep}{Center for High Energy Physics, Peking University, Beijing 100871, China}
\newcommand{\pkuphy}{School of Physics, Peking University, Beijing 100871,
	China}

\newcommand{\Uconn}{Department of Physics, University of Connecticut, Storrs, CT 06269, USA}

\newcommand{\MIT}{Center for Theoretical Physics, Massachusetts Institute of Technology, Cambridge, MA 02139, USA}

\begin{document}
	\title{Lattice QCD calculation of the subtraction function in forward Compton amplitude}
	
	\author{Yang~Fu}\affiliation{\MIT}\affiliation{\pkuphy}
	\author{Xu~Feng}
\email{xu.feng@pku.edu.cn}\affiliation{\pkuphy}\affiliation{\innovation}\affiliation{\chep}
	\author{Lu-Chang Jin}
\email{ljin.luchang@gmail.com}\affiliation{\Uconn}
	\author{Chuan~Liu}\affiliation{\pkuphy}\affiliation{\innovation}\affiliation{\chep}
	\author{Shi-Da Wen}\affiliation{\pkuphy}
	%
\preprint{MIT-CTP/5799}	

	\date{\today}
	
	\begin{abstract}
	The subtraction function plays a pivotal role in calculations involving the forward Compton amplitude, which is crucial for predicting the Lamb shift in muonic atom, 
	as well as the proton-neutron mass difference.
	In this work, we present a lattice QCD calculation of the subtraction function using two domain wall fermion gauge ensembles near the physical pion mass.
	We utilize a recently proposed subtraction point, demonstrating its advantage in mitigating statistical and systematic uncertainties by eliminating the need for ground-state subtraction. 
	Our results reveal significant contributions from $N\pi$ intermediate states to the subtraction function. Incorporating these contributions, we compute the proton, neutron and 
	nucleon isovector subtraction functions at photon momentum transfer 
	$Q^2\in[0,2]$ GeV$^2$. For the proton subtraction function, we compare our lattice results with chiral perturbation theory prediction at low $Q^2$ and with the results from the perturbative 
	operator-product expansion at high $Q^2$.
	Finally, using these subtraction functions as input, we determine their contribution to two-photon exchange effects in the Lamb shift and isovector nucleon electromagnetic self-energy.
	\end{abstract}
	
	\maketitle
	
	\section{Introduction}
	The Lamb shift has been instrumental in advancing our understanding of quantum electrodynamics (QED) and the structure of the atom. 
	It refers to a slight difference in energy levels within the hydrogen atom that was not predicted by the Dirac equation.
	Modern precision measurements of the Lamb shift have reached a point that they can probe the internal structure of the proton. A landmark example is the 2010 measurement 
	of the Lamb shift in muonic hydrogen~\cite{Pohl:2010zza}, which yielded an exceptionally accurate determination of the proton charge radius - an order of magnitude more precise than 
	previous world averages like the CODATA-2010 value~\cite{Mohr:2012tt}.
	Theoretical uncertainties in extracting the proton charge radius from the muonic hydrogen Lamb shift are mainly due to the two-photon exchange (TPE) contribution, where the non-perturbative 
	hadronic part is linked to the forward Compton amplitude.
	This presents a promising target for lattice QCD, with the first such calculation recently 
	published~\cite{Fu:2022fgh}. On the other hand,
	previous studies have largely relied on dispersive analyses using experimental data as input~\cite{Pachucki:1999zza,Martynenko:2005rc,Carlson:2011zd,Gorchtein:2013yga,Tomalak:2018uhr}, 
	supplemented by effective field theories~\cite{Nevado:2007dd,Birse:2012eb,Alarcon:2013cba,Peset:2015zga,Gasser:2015dwa,Alarcon:2020wjg,Lozano:2020qcg}, non-relativistic QED~\cite{Hill:2011wy}, and 
	the operator-product expansion (OPE)~\cite{Hill:2016bjv}.
	Such analyses require a subtraction function to ensure the convergence of the dispersive integral at large momentum transfers, and large
	uncertainties in this subtraction function limit the precision of the TPE calculations.
	A similar challenge arises in the dispersive analysis of the isovector nucleon electromagnetic self-energy, where the largest uncertainty also stems from 
	the subtraction function~\cite{Walker-Loud:2012ift,Thomas:2014dxa,Tomalak:2018dho,Gasser:2020mzy,Gasser:2020hzn}.
	
	The study of the subtraction function dates back to 1950s. In a foundational paper~\cite{Chew:1957tf}, Chew, Goldberger, Low, and Nambu developed dispersion relations for forward Compton 
	scattering, introducing a subtraction constant (the subtraction function in modern terminology) to account for uncertainties at low momentum transfers.  While much of the Compton amplitude 
	can be derived from experimental data using principles like analyticity and unitarity, the subtraction function remains inaccessible from direct measurements, making it a persistent challenge in 
	modern dispersive analyses.
	Current calculations of the subtraction function rely on models, leading to significant variation depending on the chosen approach. Efforts to constrain the subtraction function incorporate 
	theoretical insights at both low~\cite{Caswell:1985ui,Pachucki:1996zza,Pineda:2002as,Pineda:2004mx,Hill:2011wy,Carlson:2011zd,Hill:2012rh,Biloshytskyi:2023fyv} 
	and high momentum transfers~\cite{Collins:1978hi,Hill:2016bjv}, but the intermediate region remains poorly understood.
	A recent proposal suggests that dilepton electroproduction could further constrain the subtraction function, offering hope that future precision experiments at electron facilities such as MAMI, MESA, 
	and JLab might shed light on this key quantity~\cite{Pauk:2020gjv}.
	Theoretically, a precise ab initio calculation, such as from lattice QCD, would greatly enhance the accuracy of theoretical predictions. Here, we present our lattice QCD calculation of the 
	subtraction function.
	
	\section{Choice of subtraction point}
	
	In Euclidean space, the spin-averaged forward doubly-virtual Compton scattering tensor is expressed in
	\ba
	\label{eq:Compton_tensor}
		\mathcal{T}_{\mu\nu}(P,Q) &=& \frac{1}{8\pi M} \int \ud^4 x\, e^{iQ \cdot x} \langle N \vert \operatorname{T}[J_\mu(x) J_\nu(0)] \vert N \rangle 
		\nn \\
		&=&K_1(P,Q)\mathcal{T}_1(\nu, Q^2)+K_2(P,Q)\mathcal{T}_2(\nu, Q^2),
	\ea
	where
	\ba
		K_1(P,Q)& =& -\delta_{\mu \nu} + \frac{Q_\mu Q_\nu}{Q^2},
		\nn \\
		K_2(P,Q)&=&- \frac{1}{M^2}\left( P_{\mu} - \frac{P \cdot Q}{Q^2} Q_{\mu}\right)\left( P_{\nu} - \frac{P \cdot Q}{Q^2} Q_{\nu}\right),
	\ea
	and $\nu = P\cdot Q/M$, with $P= (iM,\vec{0})$ and $Q=(Q_0,\vec{Q})$ representing the Euclidean nucleon and photon four-momenta, respectively. $M$ is the nucleon mass, $J_{\mu,\nu}$ are the 
	electromagnetic quark currents and $\mathcal{T}_{1,2}(\nu,Q)$ are the Lorentz scalar functions. When needed, we use the superscripts $(\mathrm{p})$ and $(\mathrm{n})$ to 
	distinguish between proton and neutron scalar functions. The isovector Lorentz scalar functions are
	defined as $\mathcal{T}_{1,2}^{(\mathrm{v})}=\mathcal{T}_{1,2}^{(\mathrm{p})}-\mathcal{T}_{1,2}^{(\mathrm{n})}$.
	
	The TPE correction to the Lamb shift of a hydrogen-like atom is
	\ba 
	\label{eq:TPE}
		\Delta E &=&  -\frac{8 m \alpha_{\mathrm{em}}^2}{ \pi } \abs{\phi_n(0)}^2 \int \ud^4Q 
		\nn \\
		&\times& \frac{ (Q^2+2Q_0^2) \mathcal{T}_1(iQ_0,Q^2) - (Q^2 -Q_0^2) \mathcal{T}_2(iQ_0,Q^2) }
		{ Q^4 (Q^4 + 4m^2 Q_0^2) },
		\nn\\
	\ea
	with $\alpha_{\mathrm{em}}$ the fine structure constant, $m$ the lepton mass and $\abs{\phi_n(0)}^2$ the square of the $nS$-state wave function at the origin~\cite{Carlson:2011zd}.
	Similarly, $\mathcal{T}_{1,2}(\nu,Q)$ also appear in the nucleon electromagnetic self-energy, involving an integral of
	\be
	\delta M^{\gamma}=\frac{\alpha_{\mathrm{em}}}{2\pi^2}\int_R \ud^4Q\,
	\frac{-3Q^2\mathcal{T}_1(iQ_0,Q^2)+{\bf Q}^2\mathcal{T}_2(iQ_0,Q^2)}{Q^4}.
	\ee
	The subscript $R$ indicates that the integral has been renormalized, as discussed in~\cite{Walker-Loud:2012ift}.
	
	The scalar function $\mathcal{T}_{1,2}(\nu,Q^2)$ encapsulate the complex structural information of the nucleon. 
	They can be determined using dispersion relations~\cite{Hagelstein:2015egb} 
	\ba
	\label{eq:dispersion_relation}
	\mathcal{T}_1(\nu,Q^2)&-&\mathcal{T}_1(\nu_0,Q^2)
	\nn\\
	&=&\int_{\nu_{\mathrm{el}}^2}^\infty \frac{\ud{\nu'}^2}{\pi}\frac{(\nu^2-\nu_0^2)\operatorname{Im}\mathcal{T}_1(\nu',Q^2)}{({\nu'}^2-\nu^2)({\nu'}^2-\nu_0^2)},
	\nn\\
	\mathcal{T}_2(\nu,Q^2)&=&\int_{\nu_{\mathrm{el}}^2}^\infty \frac{\ud{\nu'}^2}{\pi}\frac{\operatorname{Im}\mathcal{T}_2(\nu',Q^2)}{{\nu'}^2-\nu^2},
	\ea
	where $\nu_{\mathrm{el}}=Q^2/(2M)$. The imaginary parts of $\mathcal{T}_{1,2}(\nu,Q^2)$ 
	are related to the structure functions $F_{1,2}(\nu,Q^2)$ as
	\ba
	&&\operatorname{Im}\mathcal{T}_1(\nu,Q^2)=\frac{1}{4M}F_1(\nu,Q^2),
	\nn\\
	&&\operatorname{Im}\mathcal{T}_2(\nu,Q^2)=\frac{1}{4\nu}F_2(\nu,Q^2).
	\ea 
	$F_{1,2}(\nu,Q^2)$ contain both elastic and inelastic contributions. The elastic part is described by the form factors
	\ba
	&&F_1^{\mathrm{el}}(\nu,Q^2)=\frac{1}{2}G_M^2(Q^2)\delta(1-x)
	\nn\\
	&&F_2^{\mathrm{el}}(\nu,Q^2)=\frac{G_E^2(Q^2)+\tau_p G_M^2(Q^2)}{1+\tau_p}\delta(1-x),
	\ea
	where $\tau_p=Q^2/(4M^2)$. The Bjorken variable is defined as $x\equiv Q^2/(2M\nu)$, so that $\delta(1-x)=\nu_{\mathrm{el}}\delta(\nu-\nu_{\mathrm{el}})$.
	$G_{E/M}(Q^2)$ represent the electric and magnetic form factors. For the inelastic part, $F_{1,2}^{\mathrm{inel}}(\nu,Q^2)$ are defined above the inelastic threshold, 
	$\nu_{\mathrm{inel}}=m_\pi+(m_\pi^2+Q^2)/(2M)$,
	and can be measured through electron or muon inelastic scattering.

	In the dispersive framework, $\mathcal{T}_1(\nu_0,Q^2)$ in Eq.~(\ref{eq:dispersion_relation}) represents the so-called subtraction function, 
	with $\nu_0$ being the subtraction point, which is often chosen arbitrary.
	A common choice is $\nu_0=0$, as adopted by CSSM-QCDSF-UKQCD Collaboration in their exploratory lattice QCD study using Wilson fermions, 
	which was published in the proceedings~\cite{CSSMQCDSFUKQCD:2022fzy} and has recently appeared on arXiv~\cite{Can:2025jzf}. 
	Another subtraction point, $\nu_0=\frac{i|Q|}{2}$, where $|Q|$ denotes the magnitude of the 4-vector $Q$, 
	has been proposed for the proton-neutron mass splitting~\cite{Gasser:2020mzy,Gasser:2020hzn}.
	Recently, a new subtraction point at
	$\nu_0=i|Q|$ was suggested, which has the advantage of minimizing the inelastic contribution from the structure functions $F_{1,2}(\nu,Q^2)$ to the muonic hydrogen Lamb shift~\cite{Hagelstein:2020awq}.
	
	In this lattice QCD study, we find that choosing $\nu_0=i|Q|$ provides the most accurate estimate of the subtraction function. To illustrate this, we first introduced 
	a generalized subtraction point, $\nu_0=i\xi |Q|$,
	where the four-momentum is specified as $Q_0=\xi |Q|$ and $|{\bf Q}|=\sqrt{1-\xi^2}|Q|$. Using this setup, we derive $\mathcal{T}_1(i\xi |Q|,Q^2)$ from Eq.~(\ref{eq:Compton_tensor})
	\ba
	\label{eq:subtraction_term}
	\mathcal{T}_1(i\xi |Q|,Q^2)&=&\frac{1}{2}\left[\frac{\xi^2}{1-\xi^2}\mathcal{T}_{00}-\sum_i\mathcal{T}_{ii}\right]
	\nn\\
	&=&\frac{1}{16\pi M}\int_{|t|<t_0} \ud^4x\,e^{iQ\cdot x}
	\nn\\
	&&\hspace{1cm}\times\left[\frac{\xi^2}{1-\xi^2}H_{00}(x)-\sum_i H_{ii}(x)\right],
	\ea
	where $H_{\mu\nu}(x)=\langle N|T[J_\mu(x)J_\nu(0)]|N\rangle$ and $t_0$ represents the temporal truncation. According to the current conservation and Ward identity,
	the hadronic function $H_{\mu\nu}(x)$ satisfies~\cite{Fu:2022fgh}
	\ba
	\label{eq:low_energy_expansion}
	&&\int_{|t|<t_0}\ud^4x\,H_{00}(x)=4Mt_0\,G_E^2(0),
	\nn\\
	&& \int_{|t|<t_0}\ud^4x\,\sum_iH_{ii}(x)=6\,G_E^2(0),
	\ea
	where a sufficiently large $t_0$ ensures the nucleon ground-state dominance in $H_{00}(x)\Big|_{t=t_0}$.
	These relations were also numerically confirmed in our previous study~\cite{Wang:2023omf}.

	We focus on $\mathcal{T}_1(i\xi |Q|,Q^2)$ at low $Q^2$. However, as
	$Q\to 0$, we encounter the divergence as $t_0$ increases
	\be
	\mathcal{T}_1(i\xi |Q|,Q^2)\Big|_{Q=0}=\frac{1}{8\pi M}\left[\frac{\xi^2}{1-\xi^2}2Mt_0-3\right]G_E^2(0).
	\ee
	The issue arises because the limit $Q\to 0$ is taken before $t_0\to \infty$.
	To resolve this, $t_0\to \infty$ must be taken first. Since lattice QCD data cannot access infinitely large time slices, we reconstruct the lattice data for $|t|>t_0$ using
	infinite-volume reconstruction method~\cite{Feng:2018qpx,Feng:2021zek}, or equivalently, replace the large-$t$ contribution with the ground-state contribution
	\ba
	\label{eq:subtraction_term_modified}
	\mathcal{T}_1(i\xi |Q|,Q^2)&=&\frac{1}{2}\left[\frac{\xi^2}{1-\xi^2}(\mathcal{T}_{00}-\mathcal{T}_{00}^{GS})-\sum_i(\mathcal{T}_{ii}-\mathcal{T}_{ii}^{GS})\right]
	\nn\\
	&&+\frac{1}{8\pi E}G_E^2(Q_{\mathrm{on}}^2),
	\ea
	where $Q_{\mathrm{on}}^2=2M(E-M)$ and $E=\sqrt{M^2+{\bf Q}^2}$. Here, $\mathcal{T}_{\mu\nu}^{GS}$ denotes the ground-state contribution to $\mathcal{T}_{\mu\nu}$.
	The derivation of Eq.~(\ref{eq:subtraction_term_modified}) is provided in the Supplemental Material~\cite{SM}.
	
	Although Eq.~(\ref{eq:subtraction_term_modified}) allows for calculating the subtraction function at any subtraction point, the cancellation between $\mathcal{T}_{\mu\nu}$ and $\mathcal{T}_{\mu\nu}^{GS}$
	introduces significant statistical and systematic uncertainties. However, a special case arises when $\xi=1$, or equivalently $\nu_0=i|Q|$, where the expression of $\mathcal{T}_1(i|Q|,Q^2)$ simplifies to
	\ba
	\label{eq:subtraction_term_new}
	\mathcal{T}_1(i|Q|,Q^2)&=&-\frac{1}{3}\sum_iT_{ii}
	\nn\\
	&=&-\frac{1}{24\pi M}\int \ud t\,\cos(|Q|t)H(t).
	\ea
	Here hadronic function $H(t)$ is defined as $H(t)\equiv\int \ud^3 {\bf x}\,\sum_iH_{ii}(x)$. 
	The derivation of Eq.~(\ref{eq:subtraction_term_new}) is provided in the Supplemental Material~\cite{SM}.
	Compared to the previous formalism~(\ref{eq:subtraction_term_modified}), this approach no longer requires the subtraction of the ground-state contribution. 
	From Eq.~(\ref{eq:low_energy_expansion}), we know that $\lim_{Q\to0}\mathcal{T}_1(i|Q|,Q^2)=-\frac{G_E^2(0)}{4\pi M}$.
	Thus, we define $T_1(i|Q|,Q^2)$ as
	\ba
	T_1(i|Q|,Q^2)&\equiv& -\frac{4\pi \alpha_{\mathrm{em}}}{Q^2}\left[\mathcal{T}_1(i|Q|,Q^2)-\lim_{Q\to 0}\mathcal{T}_1(i|Q|,Q^2)\right]
	\nn\\
	\label{eq:method_0}
	&=&-\frac{\alpha_{\mathrm{em}}}{6M}\int \ud t\,\frac{1-\cos(|Q|t)}{Q^2}H(t)
	\\
	\label{eq:method_1}
	&=&\frac{\alpha_{\mathrm{em}}}{6M}\int \ud t\,\frac{\cos(|Q|t)}{Q^2}H(t)+\alpha_1.
	\ea
	with $\alpha_1=-\frac{\alpha_{\mathrm{em}} G_E^2(0)}{MQ^2}$.
	Compared to $\mathcal{T}_1(i|Q|,Q^2)$, $T_1(i|Q|,Q^2)$ provides clearer insight into the low-$Q^2$ behavior. Additionally, direct insertion of $\mathcal{T}_1(i|Q|,Q^2)$
	into Eq.~(\ref{eq:TPE}) leads to infrared divergence, whereas $T_1(i|Q|,Q^2)$ is the commonly used function for calculating the Lamb shift~\cite{Hill:2011wy,Hill:2016bjv,Biloshytskyi:2023fyv}.
	Therefore, in this calculation, $T_1(i|Q|,Q^2)$ is our primary target. Once $T_1(i|Q|,Q^2)$ is determined, $\mathcal{T}_1(i|Q|,Q^2)$ is also known. 

	\section{Numerical results}
	
	We used two $2+1$-flavor domain wall fermion ensembles from the RBC-UKQCD Collaboration~\cite{RBC:2014ntl} near the physical pion mass (142.6(3) MeV for 24D and 142.9(7) MeV for 32Df~\cite{Lin:2024khg}),
	with similar spatial volumes ($L = 4.6$ fm) but different lattice spacings ($a^{-1} = 1.023(2)$ GeV and $1.378(5)$ GeV). 
	The hadronic functions $H_{\mu\nu}(x)$ are derived from the four-point correlation functions, incorporating both connected and disconnected diagrams.
	For each configuration, we generate 1024 point-source and 1024 smeared-source propagators at random spatiotemporal locations to 
	compute the correlation function $\langle \psi_p(t_f)J_\mu(x)J_\nu(y)\psi^\dagger_p(t_i)\rangle$, 
	using the random sparsening-field technique~\cite{Li:2020hbj,Detmold:2019fbk}. Here, $\psi_p$ and
	$\psi_p^\dagger$ are the proton annihilation and creation operators with zero spatial momentum. Additional details are provided in Refs.~\cite{Fu:2022fgh,Wang:2023omf,Ma:2023kfr}.
	
	As $Q\to0$, $\alpha_2\equiv \lim_{Q\to0}  T_1(i|Q|,Q^2)$ is known as
	\ba
	\alpha_2&=&-\frac{\alpha_{\mathrm{em}}}{12M}\int \ud t\, t^2 H(t)
	\nn\\
	&=&\alpha_E-\frac{\alpha_{\mathrm{em}}}{M}\left(\frac{G_E^2(0)+\kappa^2}{4M^2}+\frac{G_E(0)\langle r_E^2\rangle}{3}\right),
	\ea
	where $\alpha_E$ is the nucleon electric polarizability, $\kappa$ is the nucleon anomalous magnetic moment and $\langle r_E^2\rangle$
	is the squared charge radius. In Ref.~\cite{Wang:2023omf}, we confirmed that lattice QCD results for $\alpha_2$ are consistent with the PDG values~\cite{ParticleDataGroup:2022pth}. 
	To enhance the calculation, we can 
	determine $T_1(i|Q|,Q^2)$ by using experimental data for $\alpha_E$, $\kappa$ and $\langle r_E^2\rangle$ as input, and express $T_1(i|Q|,Q^2)$ as
	\be
	\label{eq:method_2}
	T_1(i|Q|,Q^2)=-\frac{\alpha_{\mathrm{em}}}{6M}\int \ud t\,\left[\frac{1-\cos(|Q|t)}{Q^2}-\frac{1}{2}t^2\right]H(t)+\alpha_2.
	\ee
	This expression is similar to those used for calculating hadronic vacuum polarization functions in muon $g-2$ studies~\cite{Bernecker:2011gh,Feng:2013xsa}.
	When incorporating the experimental input for $\alpha_2$, the uncertainty is predominantly driven by $\alpha_E$.

	\begin{figure}[htb]
	\centering
	\includegraphics[width=0.48\textwidth,angle=0]{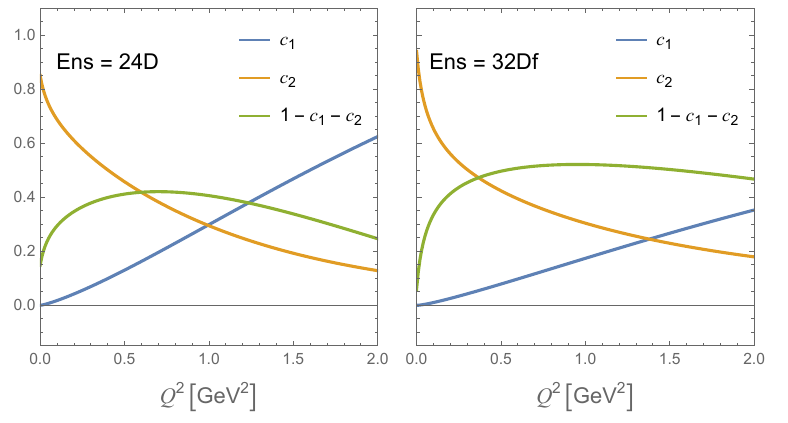}
	\caption{
	The parameters $c_1$, $c_2$ and the resulting $1-c_1-c_2$ as functions of $Q^2$ for the proton subtraction function, using both the 24D (left) and 32Df (right) ensembles.
	}
	\label{fig:c1_c2}
	\end{figure}
	
	Eq.~(\ref{eq:method_0}) is the primary formula used to calculate $T_1(i|Q|,Q^2)$, denoted as $T_1^{(0)}$. However, this method can suffer from large uncertainties at both the low- and high-$Q^2$.
	At low $Q^2$, Eq.~(\ref{eq:method_2}) provides a more accurate estimate of $T_1(i|Q|,Q^2)$ for the proton, as the uncertainty is mainly determined by $\alpha_2$, 
	which is well-constrained by experimental data.
	 At higher $Q^2$,
	Eq.~(\ref{eq:method_1}) yields better results for $T_1(i|Q|,Q^2)$ because the numerical integral is suppressed by the weight function $\cos(|Q|t)/Q^2$, and the remaining uncertainty from $\alpha_1$ is negligible.
	We define $T_1(i|Q|,Q^2)$ 
	from Eq.~(\ref{eq:method_1}) as $T_1^{(1)}$ and from Eq.~(\ref{eq:method_2}) as $T_1^{(2)}$. To take advantage of both approaches, we construct a new
	expression for $T_1(i|Q|,Q^2)$ by introducing the parameters $c_1$ and $c_2$ 
	\ba
	\label{eq:method_hybrid}
	T_1(i|Q|,Q^2)&=&(1-c_1-c_2)T_1^{(0)}+c_1T_1^{(1)}+c_2T_1^{(2)}
	\nn\\
	&=&-\frac{\alpha_{\mathrm{em}}}{6M}\int \ud t\,\left[\frac{1-c_1-\cos(|Q|t)}{Q^2}-\frac{c_2}{2}t^2\right]H(t)
	\nn\\
	&&+c_1\alpha_1+c_2\alpha_2.
	\ea
	To determine the optimal values for $c_1$ and $c_2$, we minimize the variance of the combined result $T_1(i|Q|,Q^2)$. The values of $c_1$ and $c_2$ obtained through this minimization process.
	Fig.~\ref{fig:c1_c2} shows $c_1$, $c_2$ and $1-c_1-c_2$ as functions of $Q^2$ for the case of the proton subtraction function. 	
	The corresponding results for the neutron and isovector subtraction functions are discussed in the Supplemental Material~\cite{SM}. 
	We observe similar behavior in the 24D and 32Df ensembles. 
	Note that it is not necessary to fix $c_{1,2}$  for all ensembles. In principle, $c_{1,2}$ can
	take arbitrary values. The different choices of $c_{1,2}$ may lead to varying statistical and systematic effects, but the methodology for determining $T_1(i|Q|,Q^2)$ remains consistent.
	For the proton, the PDG value for the electric polarizability is $\alpha_E^{\mathrm{(p)}}=11.2(4)\times 10^{-4}$ fm$^3$, with a few-percent precision, whereas uncertainties in the lattice results remain much larger.
	Consequently, at low $Q^2$, $c_2$ is significantly larger than both $c_1$ and $1-c_1-c_2$, indicating that $T_1(i|Q|,Q^2)$ is primarily 
	determined by $T_1^{(2)}$. As $Q^2$ increases, 
	$c_2$ decreases while $c_1$ increases, causing $T_1^{(1)}$ to dominate. However, this pattern is not universal. In the isovector case, the disconnected diagrams in the lattice calculation cancel out, 
	and the lattice results become more accurate as $Q \to 0$. Thus, the pure lattice results, $T_1^{(0)}$, dominate at low $Q^2$. Further details are available in the Supplemental Material~\cite{SM}.

	\begin{figure}[htb]
	\centering
	\includegraphics[width=0.48\textwidth,angle=0]{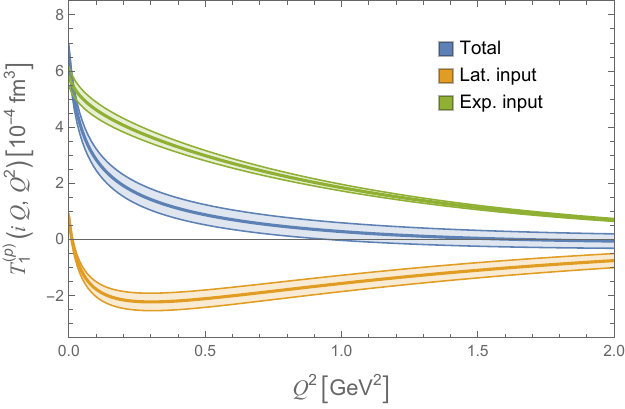}
	\caption{
	A comparison between the contributions to $T_1^{(\mathrm{p})}(i|Q|,Q^2)$ from lattice QCD input and experimental input, using the 24D proton subtraction function as an example.
	}
	\label{fig:lattice_vs_exp}
	\end{figure}

	In Eq.~(\ref{eq:method_hybrid}), the integral part
	\be
	T_1^{\mathrm{lat}}(i|Q|,Q^2)=-\frac{\alpha_{\mathrm{em}}}{6M}\int \ud t \,\omega(Q^2,t)H(t)
	\ee
	with the weight function
	\be
	\omega(Q^2,t)=\left[\frac{1-c_1-\cos(|Q|t)}{Q^2}-\frac{c_2}{2}t^2\right]
	\ee
	is influenced by the lattice input through $H(t)$, while
	the term of $c_1\alpha_1+c_2\alpha_2$ is informed by experimental data. In Fig.~\ref{fig:lattice_vs_exp}, using the 24D ensemble as an example, we distinguish these two contributions. 
	Across nearly the entire region of $Q^2\in[0,2]$ GeV$^2$, the lattice QCD input plays a significant role in determining $T_1(i|Q|,Q^2)$.
	The hadronic function $H(t)$ is subject to various systematic effects, with temporal truncation effects and finite-volume effects being the most significant in this study.
	To mitigate these systematic effects, we compute the matrix elements $\langle N|J_i(0)|N\pi\rangle$ for the four lowest $N\pi$ states in the center-of-mass frame, with nucleon momenta reaching up to
	$\sim0.5$ GeV. A detailed discussion on controlling these systematic effects can be found in the Supplemental Material~\cite{SM}.

	\begin{figure}[htb]
	\centering
	\includegraphics[width=0.48\textwidth,angle=0]{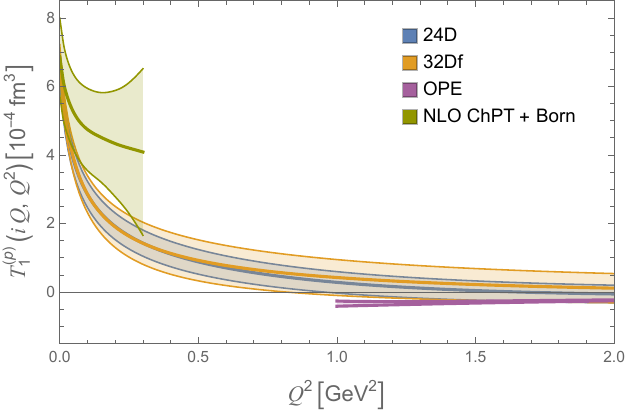}
	\caption{
	The lattice results of $T_1^{(\mathrm{p})}(i|Q|,Q^2)$ as functions of $Q^2$. These results are compared with ChPT at low $Q^2$~\cite{Biloshytskyi:2023fyv} and with the perturbative 
	OPE results at high $Q^2$~\cite{Hill:2016bjv}.
	}
	\label{fig:subtr_func}
	\end{figure}

	Fig.~\ref{fig:subtr_func} shows the final results for $T_1^{(\mathrm{p})}(i|Q|, Q^2)$, with the corresponding results for $T_1^{(\mathrm{n})}(i|Q|, Q^2)$ and $T_1^{(\mathrm{v})}(i|Q|, Q^2)$ 
	provided in the Supplemental Material~\cite{SM}. The lattice results of $T_1^{(\mathrm{p})}(i|Q|,Q^2)$ are compared with 
	chiral perturbation theory (ChPT) predictions at low $Q^2$~\cite{Biloshytskyi:2023fyv}, where the Born contribution 
	\ba
	T_1^B(i|Q|,Q^2)&=&-\frac{\alpha_{\mathrm{em}}}{MQ^2}\Big[\frac{\tau_p G_M^2(Q^2)}{1+\tau_p}+G_E^2(0)
	\nn\\
	&&-\Big(\frac{G_E(Q^2)+\tau_p G_M(Q^2)}{1+\tau_p}\Big)^2\Big]
	\ea
	has been subtracted. 
	For comparison, we add back the Born contribution using the form factors from Ref.~\cite{Borah:2020gte}. The main uncertainty in the ChPT results arises 
	from the non-Born part. At large $Q^2$, the lattice results are compared with one-loop perturbative OPE predications~\cite{Hill:2016bjv}. The results show similar behavior to both 
	ChPT at low $Q^2$ and OPE at high $Q^2$, with smaller uncertainties at low $Q^2$ and providing new insights in the intermediate region, which is not well-covered by either method.
	
	\section{Applications of subtraction function}
	Although the TPE contribution to the Lamb shift in Eq.~(\ref{eq:TPE}) is infrared divergent, this divergence can be eliminated by subtracting the contributions from a pointlike 
	term and the third Zemach moment~\cite{Carlson:2011zd}. For the subtraction-function contribution, the infrared divergence can be simply removed by performing the subtraction of 
	$\mathcal{T}_1(i|Q|,Q^2)-\lim_{Q\to0}\mathcal{T}_1(i|Q|,Q^2)$. The contribution to the Lamb shift is then given by
	\ba
	\Delta E_{\mathrm{sub}}\Big|_{\Lambda^2}&=&
	\frac{2 m \alpha_{\mathrm{em}}}{ \pi^2 }\abs{\phi_n(0)}^2
	\nn\\
	 &&\times\int_{Q^2<\Lambda^2} \frac{\ud^4Q}{Q^2}  \frac{ Q^2+2Q_0^2}
		{ Q^4 + 4m^2 Q_0^2}T_1(i|Q|,Q^2)
		\nn\\
		&=&\frac{\alpha_{\mathrm{em}}}{m}\abs{\phi_n(0)}^2\int_0^{\Lambda^2} \ud Q^2\,\frac{\gamma_1(\tau)}{\sqrt{\tau}}T_1(i|Q|,Q^2),
		\nn\\
	\ea
	where $\gamma_1(\tau)=(1-2\tau)\left[(1+\tau)^{\frac{1}{2}}-\tau^{\frac{1}{2}}\right]+\tau^{\frac{1}{2}}$ and $\tau=Q^2/(4m^2)$.
	
	By substituting the lattice results of the subtraction function $T_1(i|Q|,Q^2)$ into the integral and applying momentum cutoffs of $\Lambda^2=1$ GeV$^2$ and 2 GeV$^2$, 
	we obtain for the following subtraction-function contributions to the muonic hydrogen Lamb shift
	\ba
	\Delta E_{\mathrm{sub}}^{(\mathrm{p})}\Big|_{\Lambda^2=1~\mathrm{GeV}^2}&=&\begin{cases}
	6.94(42)(54)\mbox{ $\mu$eV} & \mbox{for 24D}\\
	7.15(70)(54)\mbox{ $\mu$eV} & \mbox{for 32Df}
	\end{cases},
	\nn\\
	\Delta E_{\mathrm{sub}}^{(\mathrm{p})}\Big|_{\Lambda^2=2~\mathrm{GeV}^2}&=&\begin{cases}
	6.97(47)(56)\mbox{ $\mu$eV} & \mbox{for 24D}\\
	7.22(81)(57)\mbox{ $\mu$eV} & \mbox{for 32Df}
	\end{cases}.	
	\ea
	The first uncertainty is the lattice statistical error, and the second is from experimental input. The results for 24D and 32Df are in very good agreement, and
	the results at $\Lambda^2=1$ GeV$^2$ and 2 GeV$^2$ are nearly the same, with only sub-percent changes. This is expected, as the integral is
	ultraviolet finite and thus not sensitive to high-momentum regions. 
	The contribution above $\Lambda^2$ can be estimated using perturbative OPE, which is found to be negligible.
	
	For neutron subtraction function, its contribution to the muonic-atom Lamb shift is
	\ba
	\Delta E_{\mathrm{sub}}^{(\mathrm{n})}\Big|_{\Lambda^2=1~\mathrm{GeV}^2}&=&\begin{cases}
	10.49(90)(77)\mbox{ $\mu$eV} & \mbox{for 24D}\\
	12.5(1.0)(1.0)\mbox{ $\mu$eV} & \mbox{for 32Df}
	\end{cases},
	\nn\\
	\Delta E_{\mathrm{sub}}^{(\mathrm{n})}\Big|_{\Lambda^2=2~\mathrm{GeV}^2}&=&\begin{cases}
	10.60(96)(80)\mbox{ $\mu$eV} & \mbox{for 24D}\\
	12.7(1.1)(1.1)\mbox{ $\mu$eV} & \mbox{for 32Df}
	\end{cases}.	
	\ea
	Here, the wave function $\abs{\phi_n(0)}^2$ is still taken from muonic hydrogen~\cite{Tomalak:2018uhr}. When applying the neutron subtraction function to muonic atoms like 
	muonic deuterium and muonic helium, $\abs{\phi_n(0)}^2$ needs to be rescaled~\cite{Krauth:2015nja,Diepold:2016cxv,Franke:2017tpc}.
	Again, the results at $\Lambda^2=1$ GeV$^2$ and 2 GeV$^2$ are consistent and the 24D and 32Df results have a slight deviation of $\sim1.5$ $\sigma$. 
	While this deviation may stem from lattice artifacts, it could also result from statistical fluctuations. Further investigation of this effect using additional 
	lattice spacings would be an interesting avenue for future study.
	
	Note that to make a meaningful comparison with previous phenomenological work, such as Ref.~\cite{Carlson:2011zd}, it is important to consider two factors: 
	(1) the results are determined using a different subtraction point, and (2) they include both the Born and non-Born contributions. 
	These differences can be accounted for and compensated by using experimental data as input.
	
	The subtraction-function contribution to the isovector nucleon self-energy can be computed as
	\ba
	\hspace{-0.5cm}\delta M_{\mathrm{sub}}^{\gamma,(\mathrm{v})}\Big|_{\Lambda^2}&=&-\frac{3\alpha_{\mathrm{em}}}{2}\int_0^{\Lambda^2}\ud Q^2\,\mathcal{T}_1^{(\mathrm{v})}(i|Q|,Q^2)
	\nn\\
	\hspace{-0.5cm}&=&\frac{3}{8\pi}\int_0^{\Lambda^2}\ud Q^2\left[Q^2T_1^{(\mathrm{v})}(i|Q|,Q^2)+\frac{\alpha_{\mathrm{em}}}{ M}\right].
	\ea
	Using the isovector subtraction function as input, we obtain the nucleon self-energy in MeV 
	\ba
	\label{eq:self_energy}
	\delta M_{\mathrm{sub}}^{\gamma,(\mathrm{v})}\Big|_{\Lambda^2=1~\mathrm{GeV}^2}&=&\begin{cases}
	-0.672(36)(04)+0.928 & \mbox{for 24D}\\
	-0.648(46)(09)+0.928 & \mbox{for 32Df}
	\end{cases},
	\nn\\
	\delta M_{\mathrm{sub}}^{\gamma,(\mathrm{v})}\Big|_{\Lambda^2=2~\mathrm{GeV}^2}&=&\begin{cases}
	-1.643(54)(09)+1.857 & \mbox{for 24D}\\
	-1.611(77)(17)+1.857& \mbox{for 32Df}
	\end{cases}.
	\nn\\
	\ea
	The two terms arise from $Q^2T_1^{(\mathrm{v})}(i|Q|,Q^2)$ and $\alpha_{\mathrm{em}}/M$, respectively. 
	The nucleon self-energy computed here is only meaningful when a momentum cutoff is applied, 
	as the integral is ultraviolet divergent. For each $\Lambda^2$, the 24D and 32Df results show good agreement. 
	By using a consistent cutoff, the lattice results can be compared with or used in the dispersive analysis, where 
	reducing model dependence is crucial for obtaining a robust error estimate.
		
	\section{Conclusion and discussion}
	In this work, we performed a detailed lattice QCD calculation of the proton, neutron, and isovector subtraction functions.
	The use of a new subtraction point allowed us to reduce statistical and systematic uncertainties, particularly by avoiding the need for ground-state subtraction. 
	The use of two gauge ensembles near the physical pion mass allowed us to obtain results very close to the physical point, but also introduced significant temporal truncation 
	and finite-volume effects. By calculating the $N\pi$-state contributions with center-of-mass momentum up to $\sim$0.5 GeV, we were able to control both of these systematic effects. 	 
	The final lattice results provide a valuable comparison with previous theoretical predictions at low and high $Q^2$, while also offering novel insights in the intermediate momentum region.
	
	We applied these subtraction functions to compute the TPE contribution to the Lamb shift in muonic atom, as well as the isovector nucleon self-energy. 
	In all cases, uncertainties were significantly reduced due to improved knowledge of the structure function in the intermediate momentum region and smaller uncertainties at low momentum.
	
	Given the challenges associated with the physical pion mass, performing the calculation at heavier pion masses, as done by the CSSM-QCDSF-UKQCD Collaboration, 
	followed by a chiral extrapolation, is a promising alternative approach. It is worth noting that even at unphysical pion masses, such as 300 MeV, controlling temporal truncation 
	and finite-volume effects remains crucial.

\begin{acknowledgments}
{\bf Acknowledgments} -- 
X.F., Y.F., C.L. and S.D.W. were supported in part by NSFC of China under Grant No. 12125501, No. 12293060, No. 12293063 and No. 12141501,
and National Key Research and Development Program of China under No. 2020YFA0406400.
Y.F. was also supported by the U.S. Department of Energy, Office of Science, Office of Nuclear Physics under grant Contract Number DE-SC0011090.
L.C.J. acknowledges support by DOE Office of Science Early Career Award No. DE-SC0021147 and DOE Award No. DE-SC0010339.
The research reported in this work was carried out using the computing facilities at Chinese National Supercomputer Center in Tianjin.
It also made use of computing and long-term storage facilities of the USQCD Collaboration, which are funded by the Office of Science of the U.S. Department of Energy.
\end{acknowledgments}

	\bibliography{ref.bib}

\clearpage

\setcounter{page}{1}
\renewcommand{\thepage}{Supplemental Material -- S\arabic{page}}
\setcounter{table}{0}
\renewcommand{\thetable}{S\,\Roman{table}}
\setcounter{equation}{0}
\renewcommand{\theequation}{S\,\arabic{equation}}
\setcounter{figure}{0}
\renewcommand{\thefigure}{S\,\arabic{figure}}

\section{Supplemental Material}

\subsection{Derivation of Eq.~(\ref{eq:subtraction_term_modified}) and (\ref{eq:subtraction_term_new})}

	If $Q$ approaches zero first, the subtraction function $T(i\xi Q,Q^2)$ defined in Eq.~(\ref{eq:subtraction_term}) encounters a linear divergence 
	as the temporal truncation parameter $t_0$ increases. To resolve this issue, we modify the expression as given in Eq.~(\ref{eq:subtraction_term_modified}). 
	Below, we provide the derivations that leads to Eq.~(\ref{eq:subtraction_term_modified}).

	The term $\mathcal{T}^{GS}_{\mu\nu}$ accounts for the ground-state contribution to $\mathcal{T}_{\mu\nu}$. 
	Before taking the limit $Q\to 0$, we first compute $\mathcal{T}^{GS}_{00}$ and $\sum_i\mathcal{T}^{GS}_{ii}$ for non-zeo $Q$, using
	\ba
	\mathcal{T}^{GS}_{\mu\nu}&=&\frac{1}{8\pi M}\int \ud^4x\,e^{iQ\cdot x}H_{\mu\nu}^{GS}(x)
	\nn\\
	&=&\frac{1}{4\pi E}\frac{Q_{\mathrm{on}}^2}{Q_{\mathrm{on}}^2+4M^2Q_0^2}\mathcal{M}_{\mu\nu}({\bf Q}),
	\ea
	where $Q=(Q_0,{\bf Q})$, $E=\sqrt{M^2+{\bf Q}^2}$, $H_{\mu\nu}^{GS}(x)$ is the ground-state contribution to $H_{\mu\nu}(x)$ and 
	\be
	\mathcal{M}_{\mu\nu}({\bf Q})\equiv \langle N(\vec{0})|J_\mu(0)|N(\vec{Q})\rangle\langle N(\vec{Q})| J_\nu(0)|N(\vec{0})\rangle.
	\ee
	The squared momentum $Q_{\mathrm{on}}^2$ results from the momentum transfer between two on-shell states and is proportional to ${\bf Q}^2$
	\be
	Q_{\mathrm{on}}^2=(iE-iM)^2+{\bf Q}^2=2M(E-M)=\frac{2M}{E+M}{\bf Q}^2.
	\ee
	By expressing $\langle N(\vec{0})|J_\mu(0)|N(\vec{Q})\rangle$ in terms of form factors, we obtain
	\ba
	T_{00}^{GS}&=&
	\frac{1}{4\pi E}\frac{4M^2{\bf Q}^2}{Q_{\mathrm{on}}^4+4M^2Q_0^2}G_E^2(Q_{\mathrm{on}}^2),
	\nn\\
	\sum_iT_{ii}^{GS}&=&-
	\frac{1}{4\pi E}\frac{Q_{\mathrm{on}}^4}{Q_{\mathrm{on}}^4+4M^2Q_0^2}
	G_E^2(Q_{\mathrm{on}}^2),
	\ea
	and their combination
	\be
	\label{eq:combination}
	\frac{1}{2}\left[\frac{\xi^2}{1-\xi^2}T_{00}^{GS}-\sum_iT_{ii}^{GS}\right]=\frac{1}{8\pi E}G_E^2(Q_{\mathrm{on}}^2).
	\ee
	By combining Eq.~(\ref{eq:combination}) with  Eq.~(\ref{eq:subtraction_term}), we immediately arrive at Eq.~(\ref{eq:subtraction_term_modified}). Since the combination of
	$T_{00}-T_{00}^{GS}$ and $T_{ii}-T_{ii}^{GS}$ completely removes the long-distance contribution, the divergence issue is resolved. Therefore, Eq.~(\ref{eq:subtraction_term_modified})
	remains valid even taking the limit $Q\to0$ before $t_0\to \infty$.

	The next task is to derive Eq.~(\ref{eq:subtraction_term_new}). Starting from Eq.~(\ref{eq:Compton_tensor}), we have
	\ba
	&&\delta_{\mu\nu}\mathcal{T}_{\mu\nu}=\mathcal{T}_{00}+\sum_i\mathcal{T}_{ii}=-3\mathcal{T}_1+(1-\xi^2)\mathcal{T}_2
	\nn\\
	&&-\frac{P_\mu P_\nu}{M^2}\mathcal{T}_{\mu\nu}=\mathcal{T}_{00}=-(1-\xi^2)\mathcal{T}_1+(1-\xi^2)^2\mathcal{T}_2
	\ea
	From the second equation, we otain
	\be
	\mathcal{T}_1=-\frac{1}{1-\xi^2}\mathcal{T}_{00}+(1-\xi^2)\mathcal{T}_2.
	\ee
	Substituting this into the first equation and imposing the condition
	$(1-\xi^2)\mathcal{T}_2\Big|_{\xi=1}=0$, we obtain
	\be
	\sum_{i}T_{ii}\Big|_{\xi=1}=\frac{2+\xi^2}{1-\xi^2}\mathcal{T}_{00}\Big|_{\xi=1}=3\frac{\mathcal{T}_{00}}{1-\xi^2}\Big|_{\xi=1}.
	\ee
	At $\xi=1$, replacing $\frac{\xi^2}{1-\xi^2}\mathcal{T}_{00}\Big|_{\xi=1}$ in Eq.~(\ref{eq:subtraction_term}) with $\frac{1}{3}\sum_{i}T_{ii}\Big|_{\xi=1}$, we arrive at Eq.~(\ref{eq:subtraction_term_new}).	
	
\subsection{Systematic effects}

	\begin{figure}[htb]
	\centering
	\includegraphics[width=0.48\textwidth,angle=0]{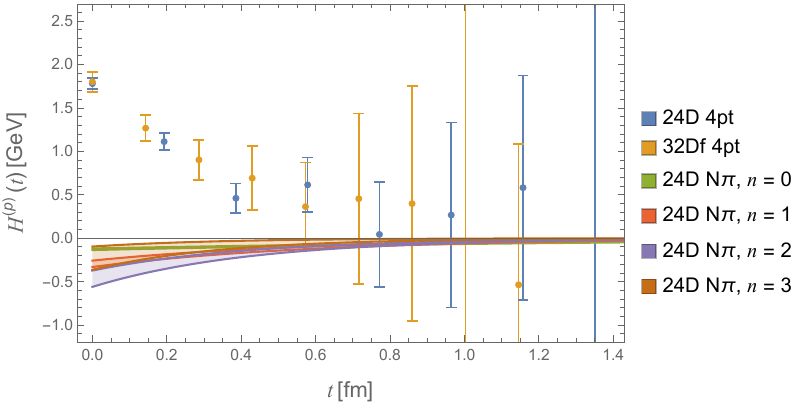}
	\caption{
	The hadronic function $H^{(\mathrm{p})}(t)$ from both the 24D and 32Df ensembles as a function of $t$. These results are compared with the contributions from the four lowest $N\pi$ states, 
	which are calculated using 24D ensemble.
	}
	\label{fig:Ht}
	\end{figure}
	
	In calculating $T_1^{\mathrm{lat}}(i|Q|,Q^2)$, the lattice data for $H(t)$ become increasingly 
	noisy as the time separation $t$ between the two currents grows, due to the signal-to-noise problem. To address this, a temporal truncation $t_0$ is introduced for the integral. 
	As shown in Fig.~{\ref{fig:Ht}}, clear signals are observed at small $t$, but quickly drown in noise. The data converge to $0$ around $0.75$ fm, so we set $t_0$ accordingly. 
	However, this truncation introduces significant systematic effects caused by $N\pi$ intermediate states, which we will correct, as discussed below.

	\begin{figure}[htb]
	\centering
	\includegraphics[width=0.48\textwidth,angle=0]{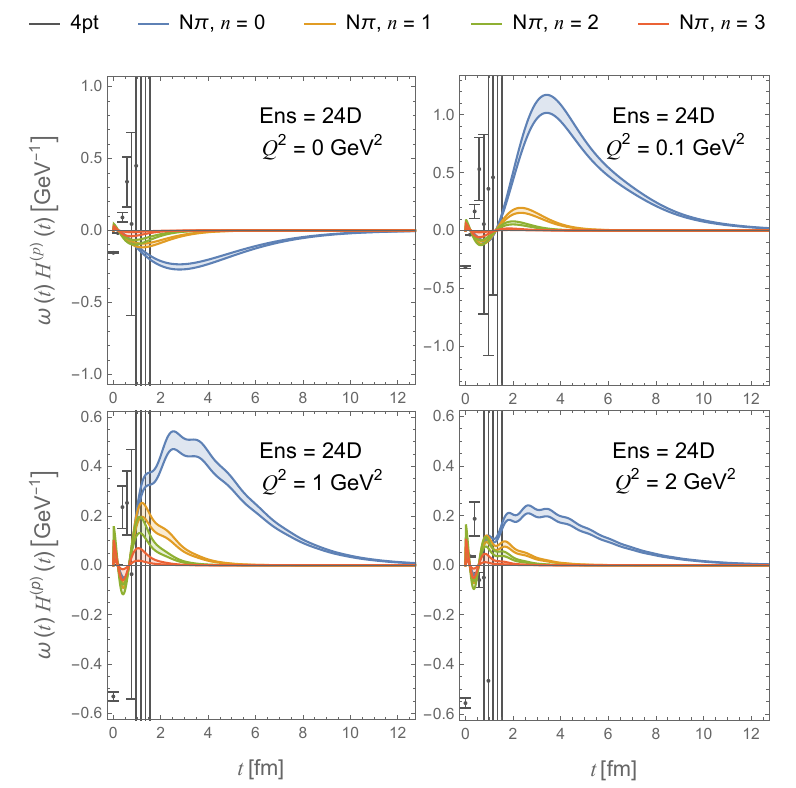}
	\caption{
	The product of the weight function $\omega(Q^2,t)$ and the hadronic function $H^{\mathrm{(p)}}(t)$ as a function of $t$ for $Q^2=0$ GeV$^2$, 
	0.1 GeV$^2$, 1 GeV$^2$ and 2 GeV$^2$. These results are compared with the contributions from the four lowest $N\pi$ states, using
	the 24D ensemble as an example.
	}
	\label{fig:weighted_Ht}
	\end{figure}

\begin{figure}[htb]
	\centering
	\includegraphics[width=0.48\textwidth,angle=0]{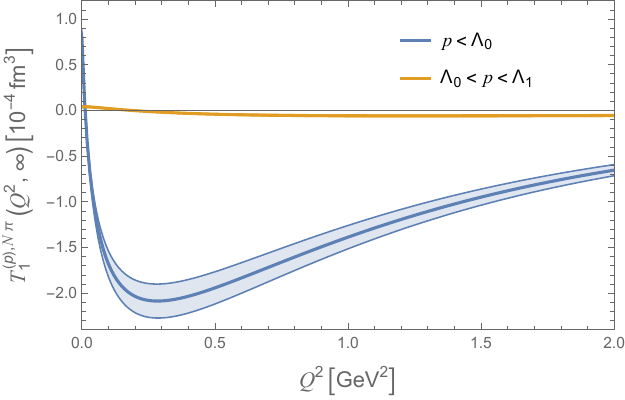}
	\caption{
	Comparison of $T_1^{N\pi}(Q^2,\infty)$ for momentum regions $|{\bf p}|<\Lambda_0$ and $\Lambda_0<|{\bf p}|<\Lambda_1$, with $\Lambda_0$ defined in 
	Eq.~(\ref{eq:mom_cutoff}) and $\Lambda_1=1$ GeV.
	The contribution from the higher momentum region $\Lambda_0<|{\bf p}|<\Lambda_1$ is substantially smaller than that from the lower momentum region $|{\bf p}|<\Lambda_0$,
	suggesting that residual temporal truncation and finite-volume effects associated with higher states are no longer significant.
	}
	\label{fig:T1Npi_p}
	\end{figure}

	At large time separation, $H(t)$ is dominated by the low-lying $N\pi$ intermediate states
	\ba
	H^{N\pi,n}(t)&=&\sum_i\langle N|J_i|N\pi,n\rangle \langle N\pi, n|J_i|N\rangle e^{-(E_{N\pi,n}-M)|t|}
	\nn\\
	&=&A_n e^{-\Delta E_n|t|},
	\ea
	where $|N\pi,n\rangle$ (with $n=0,1,2,3,\cdots$) represents the discrete energy eigenstates, ordered from lowest to highest, and $E_{N\pi,n}$ denotes their corresponding energies.
	Although $H^{N\pi,n}(t)$ are significantly smaller than the total $H(t)$, which is extracted from four-point correlation functions (as shown in Fig.~{\ref{fig:Ht}}), they are amplified by 
	the weight function $\omega(Q^2,t)$. 
	In Fig.~\ref{fig:weighted_Ht}, we demonstrate that with $t_0 = 0.75$ fm, the integral is far from converged, as the integrand associated with the lowest $N\pi$ state peaks at 
	$t_{\mathrm{peak}}=2$-4 fm for $Q^2=0$ GeV$^2$, 
	0.1 GeV$^2$, 1 GeV$^2$ and 2 GeV$^2$.
	At $Q^2=1$-2 GeV$^2$, the oscillatory behavior of $\omega(Q^2,t)H(t)$ arises from the $\cos(|Q|t)$ factor in $\omega(Q^2,t)$.
	In the previous study~\cite{Wang:2023omf}, we calculated $H^{N\pi,n}(t)$ for the four lowest states, which mitigate the temporal truncation issue. However, direct summation over the 
	discrete $N\pi$ states still suffers from significant finite-volume effects at $L\approx 4.6$ fm. To address both temporal truncation and finite-volume effects
	simultaneously, we replace the discrete finite-volume $N\pi$-state summation that includes temporal truncation at $t_0$
	\be
	T_1^{N\pi}(t_0,Q^2,L)=-\frac{\alpha_{\mathrm{em}}}{6M}\int_{|t|<t_0} \ud t\, \omega(Q^2,t)\sum_{n\le 3}A_n e^{-\Delta E_n|t|}
	\ee
	with the infinite-volume $N\pi$-state integral, which does not involve temporal truncation
	\ba
	T_1^{N\pi}(Q^2,\infty)
	&=&-\frac{\alpha_{\mathrm{em}}}{6M}\int\ud t\, \omega(Q^2,t)
	\nn\\
	&&\times\int_{|{\bf p}|<\Lambda} \frac{\ud^3 {\bf p}}{(2\pi)^3}\,A_\infty({\bf p}^2)e^{-\Delta E |t|}
	\ea
	where $\Delta E=\sqrt{M^2+{\bf p}^2}+\sqrt{M_\pi^2+{\bf p}^2}-M$, with ${\bf p}$ being the nucleon's momentum in the $N\pi$ center-of-mass frame.
	The quantity $A_\infty$ is converted from $A_n$ by multiplying a conversion factor 
	$\frac{L^3}{\nu_n}$ (with $\nu_n=1,6,12,8$ for $n=0,1,2,3$). This conversion factor is equivalent to the Lellouch-L\"uscher factor when $N\pi$-rescattering effects are neglected. 
	The ${\bf p}^2$-dependence of $A_\infty({\bf p}^2)$ has been determined in our previous study~\cite{Wang:2023omf}. 
	In relation to the discrete states $n=0,1,2,3$, we introduce the momentum cutoff $\Lambda_0$ as 
	\be
	\label{eq:mom_cutoff}
	\Lambda_0=\left(\frac{2\pi}{L}\right)\left(\left(\sum_{n=0}^3\nu_n\right)/\left(\frac{4}{3}\pi\right)\right)^\frac{1}{3}\approx0.5\mbox{ GeV}.
	\ee
	In Fig.~\ref{fig:T1Npi_p}, we compare $T_1^{N\pi}(Q^2,\infty)$ for the regions $|{\bf p}|<\Lambda_0$ and $\Lambda_0<|{\bf p}|<\Lambda_1$, where $\Lambda_1=1$ GeV. 
	We assume that ${\bf p}^2$-dependence of
	$A_\infty({\bf p}^2)$, 
	determined for $|{\bf p}|<\Lambda_0$, remains valid for $\Lambda_0<|{\bf p}|<\Lambda_1$. For $|{\bf p}|>\Lambda_1$, however, the ${\bf p}^2$-dependence of $A_\infty({\bf p}^2)$ is likely no longer applicable.
	Nevertheless, the much 
	smaller contribution from the region $\Lambda_0<|{\bf p}|<\Lambda_1$ compared to $|{\bf p}|<\Lambda_0$ suggests that the residual 
	temporal truncation and finite-volume effects associated with the higher states are no longer significant.

\subsection{Numerical results for neutron and isovector subtraction function}

	\begin{figure}[htb]
	\centering
	\includegraphics[width=0.48\textwidth,angle=0]{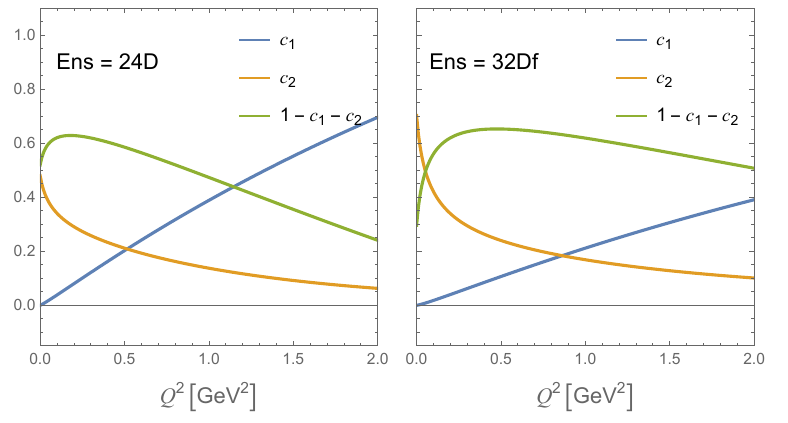}
	\caption{
	The parameters $c_1$, $c_2$ and $1-c_1-c_2$ as functions of $Q^2$ for the neutron subtraction function, using both the 24D (left) and 32Df (right) ensembles.
	}
	\label{fig:c1_c2_neutron}
	\end{figure}

\begin{figure}[htb]
	\centering
	\includegraphics[width=0.48\textwidth,angle=0]{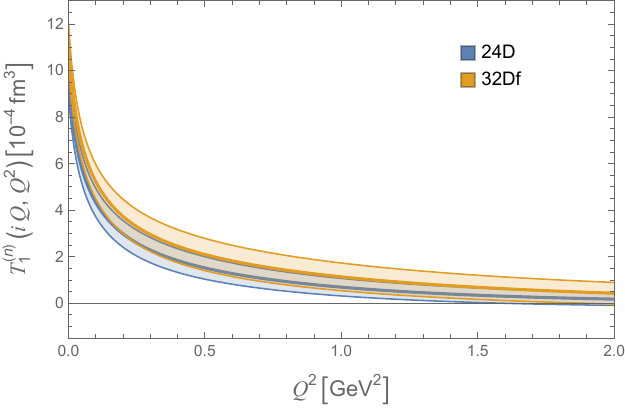}
	\caption{
	The lattice results of $T_1^{(\mathrm{n})}(i|Q|,Q^2)$ as functions of $Q^2$.
	}
	\label{fig:subtr_func_neutron}
	\end{figure}

	For the neutron subtraction function, the parameters $c_1$, $c_2$ and $1-c_1-c_2$ are shown in Fig.~\ref{fig:c1_c2_neutron}. Since the experimental measurements of $\alpha_E^{(\mathrm{n})}$ 
	are not as precise as those for $\alpha_E^{(\mathrm{p})}$, the $\alpha_2$ term does not impose a strong constraint at low $Q^2$. We observe that as $Q\to0$, the value of $c_2$ 
	deviates from 1 more significantly compared to the proton case. Using $c_1$ and $c_2$ as inputs, the final results for $T_1^{(\mathrm{n})}(i|Q|,Q^2)$ are shown in Fig.~\ref{fig:subtr_func_neutron}.
	
	\begin{figure}[htb]
	\centering
	\includegraphics[width=0.48\textwidth,angle=0]{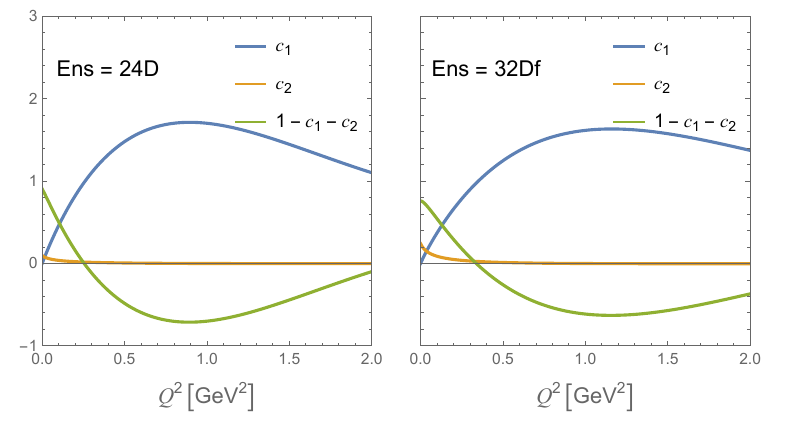}
	\caption{
	The parameters $c_1$, $c_2$ and $1-c_1-c_2$ as functions of $Q^2$ for the isovector subtraction function, using both the 24D (left) and 32Df (right) ensembles.
	}
	\label{fig:c1_c2_iso}
	\end{figure}
	
\begin{figure}[htb]
	\centering
	\includegraphics[width=0.48\textwidth,angle=0]{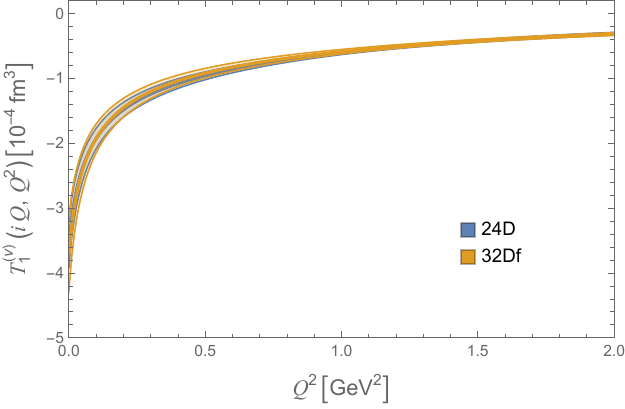}
	\caption{
	The lattice results of $T_1^{(\mathrm{v})}(i|Q|,Q^2)$ as functions of $Q^2$.
	}
	\label{fig:subtr_func_iso}
	\end{figure}

	For the isovector subtraction function, the parameters $c_1$, $c_2$ and $1-c_1-c_2$ are shown in Fig.~\ref{fig:c1_c2_iso}. We assume that the experimental measurements of $\alpha_E^{(\mathrm{p})}$ and 
	$\alpha_E^{(\mathrm{n})}$ are uncorrelated, so the combined error is taken as the square root of the sum of squares.
	In contrast, for the lattice QCD input, the disconnected diagram cancels in the isovector case, causing the lattice results to dominate the variance minimization process. 
	Consequently, $1-c_1-c_2$ is very close to 1. The final results for $T_1^{(\mathrm{v})}(i|Q|,Q^2)$ are presented in Fig.~\ref{fig:subtr_func_iso}. These results align with the combination 
	$T_1^{(\mathrm{p})}(i|Q|,Q^2) - T_1^{(\mathrm{n})}(i|Q|,Q^2)$, with only slight differences. This is because we determine $T_1^{(\mathrm{v})}(i|Q|,Q^2)$ using the $c_{1,2}$ values from 
	the variance minimization process, rather than relying on the inputs of $T_1^{(\mathrm{p})}(i|Q|,Q^2)$ and $T_1^{(\mathrm{n})}(i|Q|,Q^2)$.
	
\end{document}